\documentclass[10pt]{iopart}

\usepackage{iopams}  
\usepackage{graphicx}

\begin{document}

\title{Nonequilibrium mesoscopic transport: a genealogy}

\author{Mukunda P. Das$^1$ and Frederick Green$^2$}
\address{$^1$ Department of Theoretical Physics,
Research School of Physics and Engineering,
The Australian National University, Canberra, ACT 0200, Australia.}
\address{$^2$ School of Physics, The University of New South Wales,
Sydney, NSW 2052, Australia.}

\begin{abstract}
Models of nonequilibrium quantum transport underpin all modern
electronic devices, from the largest scales to the smallest.
Past simplifications such as coarse graining
and bulk self-averaging served well to understand electronic
materials. Such particular notions become inapplicable
at mesoscopic dimensions, edging towards the truly quantum regime.
Nevertheless a unifying thread continues to run through transport
physics, animating the design of small-scale electronic technology:
microscopic conservation and nonequilibrium dissipation.
These fundamentals are inherent in quantum transport and
gain even greater and more explicit experimental meaning in
the passage to atomic-sized devices. We review their genesis,
their theoretical context, and their governing role in the
electronic response of meso- and nanoscopic systems.
\end{abstract}

\section{Introduction}

As soon as it became possible to fabricate conducting devices at scales
comparable to the scattering mean free paths of the underlying material,
experiments began to reveal some remarkable departures from
bulk electrical-response behaviour. Among the earliest notable
phenomena were universal conductance fluctuations
\cite{lee+stone},
followed by the first measurements of conductance quantization
itself, probed as a function of carrier density in the device
\cite{delft,camb}.

These novel effects highlighted the inadequacy of standard theoretical
techniques, which had proved themselves extraordinarily effective
in the rapid development of solid-state electronics right up to the 1980s.
Typical of these older procedures is impurity-site
averaging for conductance calculations in the bulk;
relying as it does on an assumption of homogeneous randomness,
spatial averaging breaks down when the conducting channel of an actual
sample is too small to hold more than a few impurities, if any at all.

A second, more basic, instance is the limit to the applicability
of Bloch's theorem. Standard band-structure analysis fails
in the absence of lattice regularity involving, again, an
assumption of large-scale uniformity.
Few of the new types of structure are defined by
this level of periodicity.

Of prime significance to mesoscopic transport
is the openness and intimacy of contact between
the nominally active region and its macroscopic
environment. These complements -- the mesoscopic structure
and its surroundings -- are no longer subject to conventional
approximations in which one part of the complete system couples
only weakly, or quasi-independently, to the other.
The physics of the interfaces is conceptually inseparable
from the core device, and constitutes the major issue.
Even the simplest of questions, such as how to define the
very working ``length'' of such structures,
can turn into a subtle problem.

The literature is replete with texts and reviews, treating in
detail every approach to mesoscopics that now enjoys currency
\cite{imry}-\cite{12}.
In the overview to follow we aim not so much to retrace this
well-trodden ground as to probe (even at a risk of pedantry)
the fundamental underpinnings that today's mesoscopic transport theories
may wish to claim, rightly, to be integral to themselves.
For up-to-date technicalities
we direct readers to representative treatments;
those cited above are just a sample.

In meeting our goal we will cover the main conceptual points
that must underwrite successful mesoscopic approaches.
This brings us to the central question.
Precisely what are the minimal requirements on practical
methods, as imposed by the underlying principles of quantum physics?
The answer hangs upon the real nature and role of {\em multi-particle}
dynamics.

Our paper is set up as follows. In Section 2 we recall
standard methods of microscopic analysis in light of the
new problems set by carrier physics on very small scales.
The methods are: (1) the quantum Boltzmann equation;
(2) the Kubo formula; (3) field-theory based methods
(Kadanoff-Baym, Keldysh); and (4) the Landauer formula.
Despite procedural differences, the methods aim to
share a common basis in the dominance of the conservation laws.
Together with the many-body nature of transport,
that is the unifying thread.

In Section 3 we focus on models of
conductance quantization and the treatment of dissipation,
conditioned by the intimate connection with fluctuations and hence
with thermodynamic stability in any normal conduction process.
We show that exclusively linear models of mesoscopic response
are not guaranteed to capture, uniquely, the physics
of stable conduction. For, a system's stability is determined
by its essentially nonlinear dynamics; an aspect which has hardly
been investigated in mesoscopics -- and which we explore below
through a simple model. Finally we summarize this overview in Sec. 4.

\section{Approaches to Quantum Transport}

\subsection{Quantum Boltzmann Equation}

We start with some primary kinetic ideas, reaching
back to Maxwell and Boltzmann. Boltzmann's transport equation
\cite{uu,gerd,ziman1,km}
provides the opportunity to highlight concepts that have
grown to their broadest significance in the quantum realm.

A quantum extension to the Boltzmann equation,
first introduced by Uehling and Uhlenbeck
\cite{uu},
describes metallic carrier dynamics at scales of length and time
long compared to the Fermi wavelength and frequency.
A related transport equation for the degenerate,
interacting-electron liquid was formulated by Landau and Silin
\cite{dppn}.

The subject of the quantum Boltzmann equation (QBE) is the
mean distribution of carriers as it changes over time $t$ within
the single-particle configuration space for position
and wavevector $({\bf r}, {\bf k})$:

\begin{equation}
{\left[ {\partial\over \partial t}
+ {\bf v}_{\bf k}{\bf \cdot} {\partial\over \partial{\bf r}}
+ {{\bf F}({\bf r},t)\over \hbar} {\bf \cdot}
  {\partial\over \partial{\bf k}} \right]}
f_{\bf k}({\bf r}, t)
= -{\left[ {\partial f\over \partial t} \right]}_{\rm coll}
\!\!\!\!\! ({\bf k}; {\bf r}, t).
\label{n1}
\end{equation}

\noindent
With group velocity ${\bf v}_{\bf k}$ in the carrier band
and the force field ${\bf F}({\bf r},t)$ acting on each carrier,
the left-hand side of the equation gives the convective
derivative: the intrinsic or proper rate of change
of the mean particle distribution $f_{\bf k} ({\bf r}, t)$
in a frame co-moving with the semiclassical flow.
The right-hand quantity in Equation (\ref{n1}) accounts for the loss
(net outflow from the state labelled ${\bf k}$)
via the microscopic scattering processes active in the system.

The quantum Boltzmann equation provided the kinetic
template for successive, more general nonequilibrium theories.
The key to its enduring success lies in Boltzmann's original
formulation of the collision term and its method of construction.
But before anything else can be said about collision physics,
conservation must be set in place as the prime,
overarching physical constraint.

The unconditional requirement for conservation
emerges when the momentum dependence of the
one-body distribution is integrated over, to obtain the
local number density and flux; respectively,

\begin{equation}
\rho({\bf r}, t) = \Omega^{-1}\sum_{\bf k} f_{\bf k}({\bf r}, t)
~~{\rm and}~~
{\bf j}({\bf r}, t)
= \Omega^{-1}\sum_{\bf k} {\bf v}_{\bf k} f_{\bf k}({\bf r}, t)
\label{n2}
\end{equation} 

\noindent
with $\Omega$ a volume cell, centred at ${\bf r}$, over which
the band structure is (locally) defined.
Applying the trace operation of Eq. (\ref{n2})
to each side of the QBE produces

\begin{equation}
{\partial \rho\over \partial t} +
{\partial\over \partial {\bf r}} {\bf \cdot} {\bf j}
= -\Omega^{-1}\sum_{\bf k}
{\left[ {\partial f\over \partial t} \right]}_{\rm coll}
\!\!\!\!\! ({\bf k}; {\bf r}, t).
\label{n3}
\end{equation}

\noindent
In a system closed to particle exchange with
external reservoirs (as for a working electrical circuit)
the continuity equation, Eq. (\ref{n3}) above, must
be source-free. The carrier number density is microscopically
conserved if, and only if, the trace sum on the right-hand
side of the relation vanishes identically to lead to the
equation of continuity.

Why is this the crucial fact, and why even recall such an ``elementary''
given? First, it is crucial to each and every model because,
otherwise, charge could appear and disappear arbitrarily and
beyond the model's control. Nor does conservation come cheaply,
in terms of intellectual effort.
Second, no matter how cosmetically attractive
its other features might be, a scheme that was unable
to account for its microscopic conservation properties
would face a serious credibility problem, practically
as well as formally.

The primacy of conservation and continuity leaves
two options for progress. To begin with, microscopic
analysis of the Boltzmann collision term will guarantee
its vanishing trace.
Let us apply the Uehling-Uhlenbeck construction of the scattering
rate to Fermi particles
\cite{uu,ziman1}.
The term corresponding to collisions of, say, individual
electrons with a set of external scattering agents
(the particle-particle collision integral is more complex,
but analogously structured) is

\begin{eqnarray}
{\left[ {\partial f\over \partial t} \right]}_{\rm coll}
&\equiv&
\sum_{\bf k'} {\Bigl[
{\Bigl( 1 -  f_{\bf k'}({\bf r}, t) \Bigr)}
Q_{{\bf k'} \leftarrow {\bf k}} f_{\bf k}({\bf r}, t)
\Bigr.}
\cr
&& {~~~ }
{\Bigl.
- {\Bigl( 1 -  f_{\bf k}({\bf r}, t) \Bigr)}
Q_{{\bf k} \leftarrow {\bf k'}} f_{\bf k'}({\bf r}, t)
\Bigr]},
\label{n4}
\end{eqnarray}

\noindent
where the form of the transition rate
$Q_{{\bf k'} \leftarrow {\bf k}}$
carries the collision physics.

Equation (\ref{n4}) differs from the classical through its explicit
allowance for Pauli blocking of the outgoing scattering states,
a quintessentially nonclassical effect and the most overt of the
quantum phenomena entering into normal metallic transport.
The leading contribution to the right-hand
side sum sends a particle out of state ${\bf k}$ to state ${\bf k'}$.
Its counter-term accounts for the inverse process whereby a
new carrier enters state ${\bf k}$ out of ${\bf k'}$.

The scattering rate is {\em local} in space and time,
consistent with the long-wavelength time-local (Markov)
structure of the QBE's left-hand side. Other than Pauli exclusion,
all the short-range correlations or coherences in time and space,
on the scale of the Fermi frequency and wavelength, are subsumed in
$Q_{{\bf k'} \leftarrow {\bf k}}$.
These ``intra-collisional'' quantum effects do not
directly influence the QBE's kinetic shape.

The collision term on the right of Eq. (\ref{n4}) is traceless.
It satisfies conservation automatically but, in full form,
it can pose technical difficulties to an effective solution.
The level of practical effort worth spending on it
naturally depends on the fidelity we require in the answer.
Even with large computer resources the competing demands of inhomogeneity,
realistic and complex band structures, and so on may justify a
simplified approach to the scattering properties of the QBE as such.

At this point the alternative option offers itself. We can try
for a much simpler Ansatz for the collision integral.
If the new trace structure fails to vanish identically, however,
conservation may have to be imposed as an additional constraint,
no longer given for free.

An early and still popular collision approximation is the Drude model,
shown here in its most basic form:

\begin{eqnarray}
{\left[ {\partial f\over \partial t} \right]}_{\rm coll}
&\equiv&
{1\over \tau_{\rm el}}
{\left( f_{\bf k}({\bf r}, t)
- {\langle f_{\bf k'}({\bf r}, t)
\rangle}_{E_{\bf k'} = E_{\bf k}} \right)}
\cr
\cr
&& 
+ {1\over \tau_{\rm in}}
{\Bigl( f_{\bf k}({\bf r}, t) - f^{\rm eq}_{\bf k}({\bf r},t) \Bigr)}.
\label{n5}
\end{eqnarray}

\noindent
The elastic and inelastic collision times $\tau_{\rm el}$ and $\tau_{\rm in}$
reflect, in a coarse-grained way,
the more complex physics of the original scattering prescription,
Eq. (\ref{n4}).
The Drude collision times can depend on the single-particle
energy $E_{\bf k}$; here we choose $\tau_{\rm el;in}$
as constants, purely to streamline the discussion.

The equilibrium distribution $f^{\rm eq}$ within the inelastic term,
second on the right of Eq. (\ref{n5}), is
the usual Fermi-Dirac function in terms of the thermal energy $k_{\rm B}T$
and global chemical potential $\mu$, as fixed by the environment:

\begin{equation}
f^{\rm eq}_{\bf k}({\bf r}, t)
\equiv {\Bigl( 1 + \exp( (E_{\bf k}
+ u({\bf r},t) - \mu)/k_{\rm B}T ) \Bigr)}^{-1}
\label{n6}
\end{equation}

\noindent
where $u({\bf r},t)$
is an effective local potential, to be determined.
In the elastic contribution, first on the right of Eq. (\ref{n5}),
the expectation
${\langle f_{\bf k'}\rangle}_{E_{\bf k'} = E_{\bf k}}$
averages $f$ over all possible wavevector orientations,
on the assumption that an elastic collision randomizes
the exit wavevector completely, instantaneously, and
locally but without energy exchange between carrier and scatterer.

The elastic Drude term gives an instance where the
zero-trace property of the exact collision integral
is structurally preserved. Unfortunately, its inelastic partner
does not have this advantage and needs special attention.

As with the elastic component,
the assumption behind the inelastic Drude term is
that the collision instantaneously and totally randomizes the
perturbed distribution, resetting it to (local) equilibrium.
In the process the mean energy density of the carriers
is locally changed, but the particle density
{\em cannot} change. This is precisely because all scattering
processes are conceived as strictly localized
in real space. The need for conservation forecloses any option
for a discontinuous change in local carrier number.

A conserving procedure for the inelastic Drude term was
originally proposed by Greene {\em et al}.
\cite{quinn}
and independently by Mermin
\cite{mermin}.
The local potential $u({\bf r},t)$ is determined
by demanding that the trace of the inelastic term
should be zero for all $({\bf r},t)$.
For the actual system density $\rho({\bf r},t)$ this requires that

\begin{equation}
\Omega^{-1}\sum_{\bf k} f^{\rm eq}_{\bf k}({\bf r},t)
\equiv \rho({\bf r},t). 
\label{n7}
\end{equation}

\noindent
Continuity adds a consistency condition for 
$u({\bf r},t)$ as, from Eq. (\ref{n7}), it further demands that

\begin{eqnarray}
{\partial \rho\over \partial t}
&=&
\Omega^{-1}\sum_{\bf k}
{\left( -{\partial f^{\rm eq}_{\bf k}\over \partial \mu} \right)}
{\partial u\over \partial t}
= -{\partial u\over \partial t}{\partial \rho\over \partial \mu}
~\equiv~
-{\partial\over \partial {\bf r}}{\bf \cdot} {\bf j}~;
~~{\rm so}~~
\cr
\cr
{\partial u\over \partial t}
&=&
{\partial \mu\over \partial \rho}
\Omega^{-1}\sum_{\bf k} {\bf v_{\bf k}}{\bf \cdot}
{\partial\over \partial {\bf r}} f_{\bf k}({\bf r}, t).
\label{n8}
\end{eqnarray}

\noindent
Equation (\ref{n8}) defines the local time dependence of the
effective equilibrium,  subject to the model's assumption
that all relaxation events are instantaneous.

At global equilibrium we have
${\bf F}({\bf r}) \equiv -\partial u/\partial {\bf r}$
and the Fermi-Dirac distribution of Eq. (\ref{n6})
exactly satisfies the stationary, collisionless QBE
\cite{footnote1}.
Away from equilibrium, the local structure of
$f^{\rm eq}_{\bf k}({\bf r}, t)$
forces the solution to be conserving.
By a continuous {\em and local} resetting of the effective
(though not the global) chemical potential $\mu - u({\bf r},t)$,
the revised inelastic Drude Ansatz obeys the zero-trace requirement
at every step of the model system's evolution. (Physically
it also conforms to a thermodynamic rule of thumb: a disturbed
system's eventual global relaxation tends first to pass
rapidly to a temporary, locally defined equilibrium.)

With continuity now well controlled, the local equilibrium function
forms the physical input to the otherwise purely formal
solution of the full nonequilibrium distribution function.
However, there is a cost. The procedure introduces a self-consistent
loop into the calculation since, through $\rho$ and therefore $u$,
the reference equilibrium distribution is itself a functional of the
actual distribution to be solved.

Here a further consideration appears: 
regardless of its inner detail, the QBE standing alone does not
contain enough structure to close the physical transport problem.
The computational loop requires a connection
between the one-body nonequilibrium density $\rho$ and the
locally induced force field ${\bf F}$.
In the context of electron transport, closure comes through
the constitutive relation that is the Poisson equation,
tying $u$ to $\rho$ and generating the local induced electrostatic
field, adjunct to the externally supplied driving field.

Conservation has to be taken seriously. If so, it is true that
simplifying the collisional part of a transport problem is offset
by the extra effort to obtain the auxiliary distribution.
Nevertheless, as a practical matter, self-consistency cannot
be avoided. For systems with nonuniform charge distributions
(including almost every realistic mesoscopic case)
kinetics and electrostatics become inextricable. A familiar
example is the ``pinch-off'' phenomenon at the heart
of a field-effect transistor's action.

\subsection{Kubo Formula}

Kubo
\cite{kubo,ziman2,mahan}
established a very general transport framework complementary
to earlier kinetic methods. To appreciate the difference we
recapitulate the goals of each perspective.

Both in classical and quantum kinetics one starts from the abstract
but complete multi-particle description in the system's phase space,
and takes expectation values in the dynamical (Liouville) equation.
Statistical averages are first performed over the general
evolution, {\em before} any particular observable is identified
for solution, and then subsets of the internal variables are
further removed by systematic integration; a simple
illustration is the reduction of Eq. (\ref{n1}) to (\ref{n3}).

The result is an enormous nesting of reduced but
strongly coupled equations of motion for all the expectation values.
To terminate the otherwise unmanageable hierarchy,
higher-order many-body
expectations must be factorized, in some physically guided way,
and modelled as functionals of the lower-order ones. The classic
instance of this is the ``Stosszahlansatz'' of Boltzmann:
the stochastic argument leading to the characteristic shape
of his collision terms.

As Section 2.1 shows, it is essential for any such averaging to
preserve continuity, first and foremost at the single-carrier
level (analogous higher-order identities, or sum rules
\cite{dppn},
also exist).
That this is a nontrivial matter is evident in the careful
steps needed to make even the ``simple'' Drude model
microscopically conserving.

By contrast, although in Kubo's method one also starts with the system's
full quantum dynamical equation -- given the Hamiltonian -- the
quantum commutation relations are first applied to it systematically.
This has the effect of projecting the single-particle current operator,
for example, {\em uniquely} into an equivalent, two-particle
quantum correlation operator. Only after this mathematical procedure
are physical averages taken on each side of the exact relation.
This prescribes the specific relation between the mean, measurable
current response and the mean (in principle also measurable)
current-current correlation.

In summary, a kinetic equation works directly by functionally
linking expectation values of particle correlations, operating
at distinct orders in powers of the carrier density. This is possible
only by first making a (conserving) approximation for the higher-level
distributions in terms of the lower ones. For a Kubo formula, every
such correlation stays on an equal footing. Expectations strictly
come after the dynamical relationship between an observable
and its next-higher-order analogue has been singled out. Ensemble
averaging then projects both objects out of the original, complete
Hamiltonian evolution.

We now consider the eigenstates of the Hamiltonian
$H_0$ for a many-particle system of interest. An external
time-dependent perturbation $H'(t) \equiv \Theta(t) U(t) A$
begins to act at $t=0$ -- hence the step function $\Theta(t)$ -- such
that an external field $U(t)$ couples to a system observable $A$.
In charge transport, $A$ could be the density or the flux-density
operator.

A single assumption is made in the Kubo analysis: the
characteristic energy ${\langle H'(t)\rangle}$ must be
small enough (compared, say, to the free energy) to justify
a finite, and in practice linear, perturbation expansion
in powers of $H'(t)$.
For $t > 0$ the perturbation preserves a one-to-one mapping
between the basis states of $H_0$ and those of the complete
Hamiltonian $H(t) = H_0 + H'(t)$. The dynamical states remain
orthonormal but, to linear order, changes in their mutual
orientations in Hilbert space (relative to the initial basis)
can be shown to cancel exactly in the time-dependent expectation
of any observable $B$ in the Heisenberg picture:

\begin{equation}
{\langle B(t) \rangle}
\equiv 
{\left( \sum_n e^{-E_n(N)/k_{\rm B}T} \right)}^{-1}
\sum_n {\langle \psi_n | B(t) | \psi_n \rangle} 
e^{-E_n(N)/k_{\rm B}T}, 
\label{n9}
\end{equation}

\noindent
with

\begin{equation}
B(t)
\equiv
\exp{\left( {i\over \hbar} \int^t_0 H(t') dt' \right)}
\!\cdot\! B \!\cdot\! \exp{\left( -{i\over \hbar}
\int^t_0 H(t'') dt'' \right)}.
\label{n10}
\end{equation}

\noindent
The trace in Eq. (\ref{n9}) runs over the equilibrium
distribution for the whole $N$-particle system.
Only those terms directly involving the time development of $B(t)$
contribute, since the expectations themselves
are simply traces over the unperturbed
eigenstates, populated at thermal equilibrium.

The convolution expressed in
Eq. (\ref{n10}) simplifies on expanding its pair of
unitary evolution operators to linear order to arrive at

\begin{eqnarray}
B(t)
&\rightarrow&
e^{iH_0t/\hbar} \!\cdot\! {\left( 1 + {i\over \hbar}
\int^t_0 e^{-iH_0t'/\hbar} H'(t') dt' e^{iH_0t'/\hbar} \right)}
\cr
\cr
&& 
\!\cdot\! B \!\cdot\!
{\left( 1 - {i\over \hbar} \int^t_0 e^{-iH_0t'/\hbar} H'(t') dt'
e^{iH_0t'/\hbar}\right)} \!\cdot\! e^{-iH_0t/\hbar} 
\cr
\cr
&\equiv&
B_0(t) + {i\over \hbar} e^{iH_0t/\hbar} \!\cdot\! \int^t_0 dt' {\left(
{\widetilde H}'(t') \!\cdot\! B - B \!\cdot\! {\widetilde H}'(t')
\right)} \!\cdot\!e^{-iH_0t/\hbar},
{~~~ ~~~ }
\label{n11}
\end{eqnarray}


\[
\fl
{\rm where}~~
B_0(t)
\equiv
e^{iH_0t/\hbar} \!\cdot\! B \!\cdot\! e^{-iH_0t/\hbar},
~~
{\widetilde H}'(t)
\equiv e^{-iH_0t/\hbar} \!\cdot\! H'(t) \!\cdot\! e^{iH_0t/\hbar}.
\]

\noindent
Expectations are taken on both sides. Since
${\langle B_0(t) \rangle}$ keeps its original equilibrium
value (normally, this vanishes) we focus on the linear
departure of the expectation

\[
{\langle \delta B(t) \rangle} \equiv 
{\langle B(t) - B_0(t) \rangle}.
\]

\noindent
The cyclic trace property
${\langle X \!\cdot\! Y \rangle} = {\langle Y \!\cdot\! X \rangle}$
allows the exponentials in ${\widetilde H}'(t)$ to be relocated
within the right-hand side of Eq. (\ref{n11}) to yield the Kubo formula

\begin{equation}
{\langle \delta B(t) \rangle}
= {i\over \hbar} \int^t_0 dt' U(t') {\langle [A, B_0(t')] \rangle}.
\label{n12}
\end{equation}

\noindent
The formula reveals a direct and deep equivalence
between a one-body average response, such as current,
on its left-hand side and a corresponding two-body
correlation function within the integral on the right.

The significance of Eq. (\ref{n12}) is that it is universal for
normal systems (those with a stable free-energy minimum).
To make it concrete, consider a simple, circular one-dimensional
(1D) metallic loop of length $L$, threaded by an even magnetic
flux changing linearly in time. The flux induces an
electric field ${\cal E}\Theta(t)$
everywhere on the perimeter. The electromagnetic Hamiltonian
couples the vector potential $-{\cal E} t\Theta(t)$,
unique up to a conservative gauge term,
to the current operator
$I(x) = qj(x)$
\cite{footnote2}.
The expectation of the current itself
responds according to

\begin{equation}
{\langle \delta I(x,t) \rangle}
= -i{q^2{\cal E}\over \hbar} \int^t_0 \!\!dt' \!\!\int^L_0 \!\!dx'
{\langle [j_0(x',0), t' j_0(x,t')] \rangle}.
\label{n13}
\end{equation}

\noindent
from which the system conductance can be defined
(uniformity means that
${\langle I(x,t) \rangle}$ in Eq. (\ref{n13}) does not depend on $x$) as

\begin{eqnarray}
G(t)
&=&
{{\langle \delta I(t) \rangle}\over {\cal E}L} 
\equiv
{q^2\over \hbar} \int^t_0 dt' t' C(t')
~~~{\rm for}
\cr
C(t')
&\equiv&
-i\int^L_0 {dx\over L}
{\langle [j_0(x,0), j_0(0,t')] \rangle}.
\label{n14}
\end{eqnarray}

Equation (\ref{n14}) is much more general that its restricted
derivation above suggests. Moreover it carries special
implications for one-dimensional mesoscopic transport.
The time integral involving the correlator $C$
is dimensionless, and behaves in a way closely similar
to the dimensionless transmission factor of the Landauer model
for quantization of $G$ in one-dimensional ballistic conductors
(see below).
For further insight, we cite the Kubo analysis of Kamenev and Kohn
\cite{kk}
for closed driven mesoscopic circuits, including the dominant
Coulomb-screening effects expected in realistic nonuniform cases.

Another perspective on Eq. (\ref{n14}) comes from its
long-time limit. Empirically, a driven normal conductor will
always reach steady state, usually very rapidly.
Therefore $\int^{\infty}_0 t C(t) dt$ should be finite;
all transients will have fully contributed to the time integral.

There is a critical proviso. The Hamiltonian
must include one or more scattering mechanisms of a
many-body nature: carrier-phonon terms, or carrier-carrier,
or both. If $H_0$ is strictly a single-particle object with
no dynamical coupling to collective excitations,
it is straightforward to show that Eq. (\ref{n14})
will not have a well-defined asymptote unless an
infinitesimally damped, but nevertheless phenomenological,
factor $e^{-\eta t}$ is adjoined to $C(t)$.

What does this mean? If $H_0$ does {\em not}
explicitly include many-body contributions of some kind,
the long-time behaviour of the
product  ${\langle [j_0(x,0), j_0(0,t')] \rangle}$
cannot guarantee convergence to a well-defined steady state.
Stabilising it with a formal {\em ad hoc} damping factor
(in fact this is the Fermi Golden Rule in disguise) is
tantamount to mimicking damping effects that could not
otherwise emerge except by multi-particle excitations;
compare the analysis given by Hershfield
\cite{sh}.

How does natural many-body damping appear in the Kubo formula?
The time dependence of the current operator $j_0(0,t)$
leads to the spreading out, among more and more collective modes,
of its self-overlap in the correlator.
In fashionable parlance, the injected single-carrier states become
progressively ``entangled'' with the far more abundant set of
multi-particle states within the environment.

At the quantum level, therefore, propagation is not by single
carriers but by {\em entangled clusters}. Furthermore this
entanglement is as much kinematic (via Pauli exclusion
and microscopic conservation)
as it is dynamic (via interactions). It follows that many-body
arguments are necessarily implicated and that
any analysis ought not to pass them by.

The act of taking the trace average destroys the coherences that
remain encoded (albeit inexorably entangled) in the commutator.
The observable result is the actual decay of $C(t)$; and by far the
most important idea to keep in mind for this mechanism, as it
manifests in $C(t)$, is that it always represents the evolution of
dynamically correlated particle-pair states and not of autonomous
single-particle entities.
This is how conservation is maintained, while the indispensable
processes of energy loss are quantified in terms of the rate of
energy redistribution from the relatively few externally excited
modes to the exponentially more numerous multi-pair excitations
of the system.
In a real scenario, the injected energy which is first spread
out among the collective modes will be further dispersed (dissipated)
by passage to the even larger thermal environment.

For a system with effective mass $m^*$, dimensional analysis
of the correlator shows that its integral scales as

\begin{equation}
\int^{\infty}_0 t C(t) dt = {\hbar\over m^*L} n\tau
\label{n15}
\end{equation}

\noindent
in which $n = {\langle \rho \rangle}$ is the mean carrier density
and $\tau$ is a time constant that characterizes the magnitude of the
integral. Clearly, while $\tau$ is the outcome of generally very
complex kinetic events deep in the system, it quantifies the
observable attrition rate due to averaging over the ``landscape''
of all physically possible many-body excitations.
The Kubo expression comes down to

\begin{equation}
G = {q^2 n \tau\over m^*L},
\label{n16}
\end{equation}

\noindent
recognisable as the classic Drude formula
\cite{dg0}.

A clarification is in order. It is sometimes claimed that the
Kubo prescription is an artifice, a somewhat academic result restricted
to idealized systems where global charge transfer poses no issue
\cite{imry}.
The implication is that, to push it into useful form,
extra assumptions foreign to the spirit of the problem are necessary.

That is not so. Passing over the fact that practical
working circuits are electrically closed
by definition
\cite{footnote3},
the Kubo formula extends with equal rigour to any sub-system of
a normal conductor (having an absolutely stable ground state),
no matter whether the structure under study has closed or
open boundary conditions, is macroscopic or mesoscopic,
homogeneous or inhomogeneous.

It can be argued that Kubo's result is all the more powerful,
applied to open structures with charge transfer to external
reservoirs. This is because of the formula's inbuilt and
particularly stringent constraints on global, over and above
simply local, charge conservation. Interested readers are referred
to the electrodynamic analyses by Magnus and Schoenmaker
\cite{wims}
and Sols
\cite{sols}
in addition to Ref.
\cite{kk}.

\subsection{Field-theoretical (Green-function) Methods}

The analysis of response and transport reaches its most complete
form in many-body quantum kinetics, nowadays set in the formalism
of quantum fields
\cite{dppn,ziman2,mahan,kb,langreth}.
This is not the place to give more than a verbal sketch
but at least it can be said that semiclassical kinetics
(post-Boltzmann), the Kubo formalism, and much more
are subsumed in the field-theoretical point of view,
commonly also referred to as the Green-function approach.

As the single-particle distribution $f_{\rm k}({\bf r}, t)$
is the fundamental quantity of the QBE, so the one-body density
matrix

\begin{eqnarray}
\rho({\bf r}, t; {\bf r'}, t')
&\equiv&
- \Theta(t-t') {\langle \psi({\bf r}, t)\psi^{\dagger}({\bf r'}, t') \rangle}
\cr
&&
+ \Theta(t'-t) {\langle \psi^{\dagger}({\bf r'}, t')\psi({\bf r}, t) \rangle}
\label{n17}
\end{eqnarray}

\noindent
becomes the fundamental one for the dynamical equations
of quantum kinetic analysis. The expression
above has the following interpretation. The Hilbert
space of many-electron states is now extended to arbitrary
numbers of electrons, and the statistical ensemble average
${\langle \cdot\cdot\cdot \rangle}$ traces over all of them.

Acting on this greatly augmented state space, the operator
$\psi^{\dagger}({\bf r'}, t')$ ``creates'' a particle at the
space-time point $({\bf r'}, t')$ by projecting any $N$-particle
state to another with $N+1$ particles,
one of which is initially at ${\bf r'}$. In the same vein
its conjugate $\psi({\bf r}, t)$ ``annihilates'' an electron
at $({\bf r}, t)$ and sends any $N$-electron state
to some $N-1$ state with an electron initially absent
at ${\bf r}$. The operators are cast in the Heisenberg picture,
as with Eq. (\ref{n10}) above. Fermion
antisymmetry means that, at equal times,
they satisfy the anticommutation relation

\[
  \psi({\bf r}, t) \psi^{\dagger}({\bf r'}, t)
+ \psi^{\dagger}({\bf r'}, t) \psi({\bf r}, t)
= \delta({\bf r} - {\bf r'})
\]

In Eq. (\ref{n17}), the time ordering
reverses the role and sign of its fermion operator pair:
for $t > t'$ we add a single-electron excitation, later to be removed.
This probes the overlap, or correlation, within the evolving
one-body distribution and is the simplest measure (though
theoretically insufficient for conductance) of the system's
relaxation during the interval $t - t'$.
Conversely, for $t' > t$ the operation removes an
electron -- adds a hole -- and restores it later
to measure relaxation of the hole in the interacting-electron background.
(The choice of time ordering is for convenience since,
mathematically, a hole that evolves forwards in time looks just
like an electron evolving ``backwards'' in time.)

Given a system's interacting Hamiltonian, its many-body
Schr\"odinger equation leads to a conjugate pair of coupled
equations of motion for $\rho({\bf r}, t; {\bf r'}, t')$.
The objects $\psi$ and $\psi^{\dagger}$ act like quantum-field
operators and $\rho$ has all the characteristics of the
Green function, or ``propagator'', for the coupled equations.
This makes the language highly abstract; in compensation
it also makes accessible, to practical condensed-matter transport
calculations, some sophisticated techniques of field theory
such as Feynman-diagram analysis
\cite{ziman2}
and Keldysh time-evolution contours
\cite{langreth}.

As a two-point double-time quantity, the complexity of the
density matrix exceeds that of the solution to the QBE.
On the other hand, it necessarily encodes far more information:
its evolution admits every possible quantum coherence effect,
at every scale. Nevertheless, despite the quantum subtleties,
the way in which the evolutionary character of $\rho$ is interpreted
stays conceptually quite close to the older understanding
that motivates the Boltzmann equation.

The quantum many-body equations' kinematic complexity is only the
beginning. The creation-annihilation pair within the density
matrix couples to the Hamiltonian such that two-pair, three-pair
and, in general, multi-pair excitations all enter and contribute to
its space-time development. Their interaction within the
density matrix appears as a hierarchy of convolutions which
represent the full generalization of Boltzmann's collision terms:
there is a ``scattering-out'' component, or self-energy, that
removes particle strength from $\rho$ and a ``scattering-in''
component, or vertex, that tends to restore particle strength.

Self-energy and vertex parts are intricate
functionals of the higher-order multi-pair excitations
coupled to the original electron-hole pair
($\psi^{\dagger}\psi$ or $\psi\psi^{\dagger}$)
represented by $\rho$. To satisfy particle conservation
it is not only imperative for both parts to be retained
in any approximate treatment of the dynamics but that they
should be inter-related in a highly specific way, through
the Ward identities
\cite{mahan}.
Strategies for securing the essential conserving
properties within approximate models of quantum transport
are fully explored by Kadanoff and Baym
\cite{kb}.
Exactly the same considerations apply to the problem
when posed in the language of Keldysh
\cite{langreth}.

The Ward identities generalize the
optical theorem of scattering
\cite{mahan},
which asserts that the loss of energy and momentum
from an initial freely propagating state must be made up
by the total gain of all the other states in the Schr\"odinger
problem. Since unitary scattering couples all such modes,
their evolution must be solved jointly.
In transport, such modes are physically occupied, so
energy-momentum conservation becomes inherently a many-body
problem. The Ward identities quantify this collective
coupling subject to conservation, in a way similar to
the optical theorem's requirement on the underlying basis
states. But while knowledge of the single-particle modes
is a prerequisite to set up the many-body transport solution,
it can never supplant it.

If the intrinsic particle-particle interactions are strong
and/or long-ranged, normal perturbation theory will not work.
The literature describes countless ingenious ways to deal with
the internal correlations. It has a long pedigree, it is vast
and it is still growing. For this discussion,
the general References 
\cite{dppn,ziman2,mahan,kb,langreth}
must suffice.

Although internal interactions may be strong, one can still
use linear response for the conductance of a metallic
electron system, if driven by a weak enough
external field
\cite{footnote4}.
Under this condition, the Kubo formalism is replicated within
the Green-function picture
\cite{mahan}.

A completely quantum treatment of $\rho$ is needed
whenever the range of its quantum correlations begins
to be reached by the physical size of the system or by
the time scale for effects of interest; in less extreme applications
the full description can be simplified. For example, to study
processes whose rates are slow compared to the inverse Fermi frequency
we can restrict the analysis to the density-matrix form

\[
\phi({\bf r}, {\bf r'}; t) \equiv
\lim_{t' \to t^+} \rho({\bf r}, t; {\bf  r'}, t')
= {\langle \psi^{\dagger}({\bf r'}, t) \psi({\bf  r}, t) \rangle}.
\]

\noindent
Electron correlations are regarded as instantaneous, while their spatial
nonlocality, or off-diagonal structure, remains an essential feature.
The overall time evolution of the system, analysed quasi-statically (relative
to the electrons), recalls the Born-Oppenheimer decoupling of
slow and fast motions within electronic systems.
The dynamics of this limit has been studied in depth by Mahan
\cite{mahan}.

The equal-time matrix $\phi({\bf r}, {\bf r'}; t)$
is governed by a so-called master equation,
one version of which is the Lindblad equation
\cite{raja}.
Its form can be distilled from its original quantum dynamical setting,
though not without additional assumptions extending
the arguments of Uehling and Uhlenbeck to deal with the off-diagonal
nature of the density matrix.

As a final step the quantum Boltzmann equation re-emerges on
restricting attention to $\phi$ in the neighbourhood of its
diagonal elements $\phi({\bf R}, {\bf R}; t)$.
From $\phi$, Wigner's distribution function can be defined as

\[
f_{\bf k}({\bf R}, t)
\equiv
\int d^3r e^{-i{\bf k}\cdot{\bf r}}
\phi({\bf R} + {\bf r}, {\bf R} - {\bf  r}, t).
\]

\noindent
If $\phi$ decreases sharply away from
the central co-ordinate ${\bf R}$, the Wigner function behaves
as a localized slowly varying quantity. Any remnant
off-diagonal nonlocality is taken up
into the dependence on wavevector ${\bf k}$. The latter
spans a background electronic band structure,
defined over some cell $\Omega$ centred on ${\bf R}$,
whose size is small on the typical scale of $|{\bf R}|$
yet still large compared to the Fermi wavelength of
the electrons; if not, the band-structure picture would
not make good physical sense. As derived by Kadanoff and Baym
\cite{kb}
(with some assumptions for the ${\bf k}$-dependent
self-energy and vertex structures destined to become
the locally defined Boltzmann collision term)
this limit of the Wigner function $f$ becomes the
semiclassical fermion occupancy,
obeying a reduced equation identical to the QBE.

Distinct from the long-wavelength recovery of semiclassical kinetics,
the original Wigner distribution can be studied and solved
as a fully quantum distribution in its own right,
equivalent to the equal-time density matrix and
with its master equation transcribed to the Fourier domain.
Wigner-function methods form an extensive body of work,
particularly in high-field problems where Kubo and linear-response
approaches lose their immediate utility.
An instructive reference to this
different transport methodology is Frensley
\cite{frensley}.

\subsection{Landauer Formula}

Anticipating the eventual creation of structures
with feature sizes much below any bulk mean free path, Landauer
\cite{rolf}
and subsequent developers of his approach
\cite{imry,markus,mw}
were motivated to understand conductance as a mesoscopic
phenomenon with less emphasis on kinetic relaxation and more on a
hydrodynamic (in the quantum limit, wave-like) concept of the
interaction with scattering sources. The sources are treated
not so much as mediators of dynamical many-body excitations
but as modifiers of single-particle ballistic ``flow'' which
remains, at all events, microscopically free and undisturbed.

The analogy shifts from the staccato motion of colliding
billiard balls towards the smooth geometry of a working airfoil.
One might say -- with a bit of poetic licence -- that the Landauer
approach to mesoscopics aims to be holistic while standard ones
are bound to remain thoroughly reductionist.
In stressing a continuum-like rather than particulate formal
description of transport, one feature of this viewpoint is a
more diffuse theoretical link between the average response of
a system and fluctuations about the average.

Landauer characterized his own viewpoint in part as ascribing
conceptual parity to the roles of current and voltage
\cite{rolf}.
Either should equally well be able to act as the causative agent
of transport, with its partner representing the effect. This is
radically different from the Kubo analysis, for example, where
an external field has absolute meaning as the applied stimulus
to conduction while the current is invariably the system's
{\em causal} response. Causality turns out to be intimately tied
to the notion of dissipative relaxation.

It may appear ironic that the Landauer formula applied to 1D
conductance does not actually require any novel assumptions,
as introduced originally by Landauer, to establish it.
To show that it does not, we derive it straightforwardly from
canonical transport theory. The proof has several stages.

\begin{itemize}
\item {\em First}: we invoke the Kubo formula, Eq. (\ref{n14}),
for the current traversing an open one-dimensional mesoscopic channel
\cite{sols};
the working channel is a uniform wire smoothly embedded within
a sub-system itself forming a segment of a macroscopically
closed driven loop
\cite{kk}.
The sub-system encompasses the mesoscopic interface regions,
since these are comparable in size to the channel's conductive core.
Even if the core is ballistic, scattering in the boundaries
is a real and dominant effect. Equation (\ref{n15}) therefore applies.

\item {\em Second}: existence of a steady state implies
that the time integral of the correlator in Eq. (\ref{n15})
is finite. Therefore the conductance takes the form of
Eq. (\ref{n16}):

\begin{equation}
G = {q^2n \over m^* L} {\tau_{\rm in}\tau_{\rm el}\over
{\tau_{\rm in} + \tau_{\rm el}}}
\label{n20}
\end{equation}

\noindent
where, by Matthiessen's rule, the stochastically independent
collision rates $\tau_{\rm in}^{-1}$ and  $\tau_{\rm el}^{-1}$ add
together and lead to the composite collision time
$\tau = (\tau_{\rm in}^{-1} + \tau_{\rm el}^{-1})^{-1}$.

\item {\em Third}: the sub-system's operative length $L$
must be determined by the physical environment of
the open configuration for the core plus boundary regions.
In a ballistic structure, the effective mean free paths are
no longer uniquely set by the bulk material,
but by the dynamics at the interfaces. This implies
that $L$ becomes, in effect, the maximum of the mean free paths
$\lambda_{\rm in}$ and $\lambda_{\rm el}$, or any others
that can be theoretically (and experimentally) distinguished over
the span of the device.

Suppose the mean speed of the carriers within the system is $v_0$. Then 

\[
L = {\rm max} {\{ \lambda_{\rm in}, \lambda_{\rm el} \}}
\equiv v_0  {\rm max} {\{ \tau_{\rm in}, \tau_{\rm el} \}}
\]

\vskip -3mm
\noindent
so
\vskip -3mm

\begin{eqnarray}
G
&=&
{q^2n \over m^* L} {\lambda _{\rm in}\lambda_{\rm el}\over
v_0(\lambda_{\rm in} + \lambda_{\rm el})}
\cr
\cr
&=&
{q^2n \over m^* v_0} {\lambda _{\rm in}\lambda_{\rm el}\over
{\rm max} {\{ \lambda_{\rm in}, \lambda_{\rm el} \}}
(\lambda_{\rm in} + \lambda_{\rm el})}
\cr
\cr
&=&
{q^2n \over m^* v_0} {\left( {1\over 2} -
{ |\lambda_{\rm in}- \lambda_{\rm el} |\over
 2(\lambda_{\rm in} + \lambda_{\rm el}) } 
\right)}.
\label{n22}
\end{eqnarray}

\item {\em Fourth}: in the degenerate limit, the characteristic
speed of the carriers becomes the Fermi velocity; $v_0 = v_{\rm F}$.
At the same time the density is given by $n = 2k_{\rm F}/\pi$
where $k_{\rm F}$ is the Fermi wavevector, related to $v_{\rm F}$
through the momentum: $\hbar k_{\rm F} = m^* v_{\rm F}$. Substitution
for $n$ and $v_0$ in Eq. (\ref{n22}) gives

\begin{equation}
G = G_0 {\left(
1 - {|\lambda_{\rm in} - \lambda_{\rm el}|\over
    {\lambda_{\rm in} + \lambda_{\rm el}} }
\right)};
~~G_0 \equiv {q^2\over \pi\hbar}.
\label{n23}
\end{equation}

\end{itemize}

\noindent
The expression in parentheses on the right-hand side of
Eq. (\ref{n23}) operates in every way identically to the
transmission factor ${\cal T}$ in the standard form
of the Landauer conductance,

\begin{equation}
G = G_0 {\cal T} ~~{\rm with}~~
~~ 0 \leq {\cal T} \leq 1.
\label{n24}
\end{equation} 


What is the experimental distinction
between Eq. (\ref{n23}) derived directly from the
standard quantum Kubo formula, and Eq. (\ref{n24})
obtained by rather different physical arguments
\cite{imry,markus}?
There is no empirical distinction at the level of
so-called ``two-probe'' current-voltage measurements
(both quantities being read off the same pair of probe terminals).

However, the underlying differences in origin and
interpretation may be substantial.
Bringing these out could be a matter of some interest,
calling for additional measurements designed around,
say, well constructed ``four-probe'' configurations
to obtain current and voltage reliably and independently
across the inner core.
Also necessary would be improved experimental control
of inelastic, over against elastic, scattering effects.

Another central quantity for testing issues of principle
is current noise, providing knowledge of the two-body correlator
$C$; recall Eq. (\ref{n14}).
Noise properties lie beyond the capacity
of two- and four-probe methods,
which test only the single-carrier response; but
high-quality nonequilibrium noise measurements are difficult
to carry through. The basic implications of existing
ballistic-noise experiments, analysed quantum kinetically,
have been addressed elsewhere
\cite{gtd}.

In the Landauer equation (\ref{n24}), the ideal scenario
plays out when $G$ reaches the full conductance quantum $G_0$.
This occurs if, and only if, ${\cal T} = 1$. The channel
is then considered to be perfectly transparent to the carriers.

The Kubo-derived Eq. (\ref{n23}) attains ideality
if, and only if, the mean free paths are {\em equally matched},
both consequently being equal to the ballistic length of the
structure as a whole: interfaces as well as interposed conductor.
This does not entail ``perfect'' -- totally collisionless -- transport
since both inelastic and elastic scattering act unfettered
at the scale of the overall structure.

It is because the mesoscopic Kubo description necessarily
incorporates the interface physics into that of the ``real'' device,
sandwiched between, that collision events retain their explicit
and pre-eminent role even in the case of ideally quantized
ballistic conductance. This differs from the Landauer
description in that (a) elastic scattering is present and robust,
and (b) the physics of dissipation does not devolve to
the asymptotic equilibrium regions in the leads
\cite{imry}
but acts co-cooperatively with the elastic effects.
In Kubo, dissipation is the inalienable signature
of dynamics in the active, {\em nonequilibrium} region.

The boundaries are always with us.
The mesoscopic boundaries are the physical loci of all scattering,
elastic and inelastic together.
Yet it is inelastic scattering alone that stabilizes the steady state for
every transport problem, irrespective of the device scale.
The deeper reason why that is so concerns us in the following
Section.

\section{Dissipation and Conductance}

\subsection{Dissipation}

As long as a conducting structure is ohmic, it dissipates
its externally supplied power at the usual rate

\begin{equation}
P = IV = G V^2.
\label{n25}
\end{equation}

\noindent
A circuit in which such dissipation occurs is insensitive
to the microscopic details of conduction; it is essential only
that there {\em exist} a path -- any path at all -- for energy loss.
Should it happen that $G = G_0 = 77.4809~\mu{\rm S}$,
it makes no difference to Eq. (\ref{n25}) whether this conductance
belongs to a common-stock resistor, or is measured for a uniquely
hand-crafted and very costly ballistic research device.

The key question is straightforward. In terms of its energetics,
what distinguishes the process of conduction in a conventional
channel from a ballistic one? A crude answer might be: nothing.

Unsatisfactory as it seems, such an answer does not fundamentally
require too much elaboration. The universality of electrical
energy dissipation is plain; what is at times less obvious is
how to resolve the apparent incompatibility between ideally ballistic
transport on the one hand and, on the other hand, the absolute
necessity for accelerated electrical charges to shed their energy
gain {\em inelastically}.

In the previous Sec. 2.2 we intimated that the physical
answer to the origin of energy dissipation rests in the correlation
structure of the Kubo formula, and in Sec. 2.4 we made direct use of
that relation to recover, in a standard way, quantization of $G$
in a one-dimensional metallic mesoscopic conductor. Commensurability
of the nondissipative (elastic) scattering length and the dissipative
(inelastic) length, and of each of them with the operational size of
the ballistic conductor, is the essential element in the emergence of
quantization within the Landauer formula, as a corollary to the
general Kubo analysis.

Dissipation is thus a given in normal transport and the necessity for
its presence is widely appreciated. Nevertheless discussion sometimes
still surrounds not the reality of dissipation, but its actual centre
of operation.
Is energy loss an active process, immediately tied to the dynamics of
the structure under test and thus demanding explicit modelling
\cite{dgvpt}?
Or is it instead a remote and passive effect, buried deep within
the equilibrium background and hence free of all need to
represent it within the physics of a viable transport model
\cite{imry}?

Kubo's formula reveals the first of the alternatives above
to be the one that faithfully reflects the physics of dissipation.
Conditioned as it is by the many-body correlation properties,
the Kubo formula demands that inelastic relaxation be present
and accounted for {\em quantitatively} within the
correlator. Were elastic scattering to be the only process allowed,
the system could not relax; it would have no mechanism for shedding
the energy pumped into its carrier population by the driving field.
The system would become grossly disequilibrated and steady state
would be impossible.

\subsection{Fluctuations}

A concrete example illustrates the drastic thermodynamic consequences
if mesoscopic transport were exclusively determined by elastic
(nondissipative) scattering. We preface the example by introducing
an essential idea, the fluctuations in a mesoscopic conductor.
To that end we recall the Kubo relation in its alternative form
\cite{ziman2,mahan}
(we will specialize to 1D),

\begin{equation}
G = {1\over k_{\rm B}T} {\langle {\cal I}_0 I_0(0) \rangle};
~~{\cal I}_0 \equiv \int^{\infty}_0 dt I_0(t).
\label{n26}
\end{equation}

\noindent
Equation (\ref{n26}) is the zero-frequency -- that is,
steady-state -- version of the fluctuation-dissipation theorem (FDT).
It asserts that the low-field conductance is defined by the
equilibrium fluctuations, or statistical self-correlation, of the
current about its zero mean; the very heart of the Kubo formula.
In Ziman's words
\cite{ziman2}:
``conductivity is a property inherent in the quantum-mechanical
description of the unperturbed system; the application of a weak
electric field merely exposes the time-correlations of the fluctuating
components of the electric current in the equilibrium state.''
This is a general reformulation of Einstein's early conclusion
that random (Brownian) motion determines the measurable mobility of
a test particle.

The fluctuation-dissipation theorem's significance is that
{\em every microscopically conserving transport model}
must imply it.
This covers every model that can be consistently
formulated within quantum kinetics (Sec. 2.3)
including, apart from the Kubo formula,
the entire class of quantum Boltzmann theories (Sec. 2.1).

We emphasize that the FDT is a necessary consequence of
conservation. A transport model, if forced to assume the
fluctuation-dissipation relation as an additional
external hypothesis, could not have incorporated
the physically required level of conservation in the first place.
As with the Drude model of Eq. (\ref{n5})
the remedy would be to restructure the model's equation of motion
to make it microscopically conserving; the theorem would then follow.
It is not enough simply to add the FDT phenomenologically,
without reconsidering the model's underlying kinetics.

The scope of the FDT omits nonequilibrium fluctuations, however.
While the latter must conform to the
FDT in the linear weak-field limit, the relation by itself cannot
reveal anything of the system's behaviour as soon as it departs
from linear response. A genuinely nonlinear kinetic framework is needed.

To compare the wider consequences of active dissipation
with those of a dissipation-free picture of ballistic conductance,
we revisit a conserving Boltzmann-Drude description of uniform
1D mesoscopic transport
\cite{fnl}.
The model recovers the FDT in the weak-field regime and
reproduces the Landauer conductance from Eq. (\ref{n23}).
Away from that limit it gives a quantitative account of
nonequilibrium current fluctuations for a
ballistic conductor at high driving fields
\cite{gtd}
and thus offers a test-bed for such effects.

We return to the zero-frequency conductance given by that model for
metallic (degenerate) carriers, but with a new working premise:
the mesoscopic sample length $L$ is to be set independently
of the scattering mechanisms at the sample boundaries
(as just one instance, it might be visualized to be
the wire's dimension in fabrication).
Combining Eqs. (\ref{n20}) and (\ref{n26}), the low-field
conductance is accompanied by the current-fluctuation strength

\[
S(0) \equiv {\langle {\cal I}_0 I_0(0) \rangle}
= k_{\rm B} T G_0
{2\lambda_{\rm in}\lambda_{\rm el}\over L(\lambda_{\rm in}+\lambda_{\rm el})}
\]

\noindent
in the zero-voltage limit.

The same theory also yields the corresponding nonlinear
current-current fluctuation at higher applied voltages.
For finite $V$ this is
\cite{fnl}

\[
S(V) = S(0) {\left[ 1 + {\left( {qV\over m^*v^2_{\rm F}} \right)}^2
{\left( {\lambda_{\rm in}\over L} \right)}^2
{\left( 2 - {\lambda_{\rm in}^2\over (\lambda_{\rm in}+\lambda_{\rm el})^2}
\right)}
\right]}
\]

\noindent
while the power dissipation in the model remains as

\[
P = V^2 G_0
{2\lambda_{\rm in}\lambda_{\rm el}\over L(\lambda_{\rm in}+\lambda_{\rm el})}
\]

\noindent
so the energy loss rate
stays directly proportional to $S(0)$.

The zero-field fluctuation strength $S(0)$ scales with the typical
energy $k_{\rm B}T$ of a carrier at thermal equilibrium.
What, then, does the difference of the current fluctuations,
$S(V) - S(0)$, represent? It can be characterized
in terms of an effective ``hot-carrier'' excess temperature
$\Delta T \equiv T(S(V)/S(0) - 1)$; that is, it measures
the overhead of energy converted into -- and retained as -- the
mean kinetic energy of highly excited carriers accelerated
by the driving field.

Thus the nature of $S(V) - S(0)$ is nondissipative; excess
fluctuations are not associated with the energy {\em transfer}
that manifests as Joule heating in the structure and its surroundings,
since this is keyed to $S(0)$. Rather, it reflects the average
inventory of energy built up and stored in the carrier
population by the underlying dynamical processes that
sustain the nonequilibrium carriers in steady state.

Consider next what would happen in the limit of asymptotically weak
inelastic scattering
\cite{footnote5};
all along, $L$ stays fixed by hypothesis.
Then $\lambda_{\rm in} \gg L, \lambda_{\rm el}$ and leads to

\begin{eqnarray}
S(0) &\to& k_{\rm B} T G_0 {2\lambda_{\rm el}\over L},
~~{\rm while}
\cr
\cr
S(V)
&\to& S(0){\left[ 1 + {\left( {qV\over m^*v^2_{\rm F}} \right)}^2
{\lambda^2_{\rm in}\over L^2} \right]}
~~{\rm so}
\cr
\cr
\Delta T
&\gg&
T.
\label{n29}
\end{eqnarray}

\noindent
This example highlights the consequence of taking inelastic
scattering completely out of the mesoscopic transport picture.
Even though conductance is not qualitatively affected,
the nonequilibrium current fluctuations diverge;
they do so, in fact, for any nontrivial applied voltage.
As a result, while the low-field current response $I = GV$ may well
remain finite -- in a strictly formal sense -- its deviation from
that average blows out uncontrollably. This makes the very possibility
of measuring the current's mean value empirically void.

A mesoscopic calculation done strictly in linear response but excluding
from consideration all inelastic collisions will, notwithstanding,
yield a plausible value for the low-field conductance -- and fail
to reveal the uncontrollably wild behaviour of the nonequilibrium current
fluctuations, which is inevitable except at $V=0$. Deprived of any
mechanism for inelastic energy loss, a prediction that appears quite
reasonable at linear response must, in reality, hide the seriously
unphysical nature of its associated fluctuations.

One can contrast the nonequilibrium implications of the calculation for
a sample length that is in fact physically delimited by the
scattering physics in the interface regions, as described previously.
With this understanding of $L$, and assuming now that the
inelastic mean free path is largest one in the device configuration,
we have $L = \lambda_{\rm in} \gg \lambda_{\rm el}$.
Substitution into Eq. (\ref{n29}) shows that $S(0)$ and
$G$ are diffusively dominated by elastic scattering. Meanwhile,
the excess fluctuations, no longer divergent, are well behaved
in all circumstances.

The example above entails four considerations:

\smallskip
(i) A description of transport in a normal, thermodynamically
stable system should incorporate a specific model for inelastic
scattering; otherwise, the existence of a meaningful steady state
is not guaranteed. This constraint applies just as much to mesoscopic
charge transport as to any other setting for electrical conduction.
Inelasticity is intimately linked to the dissipative energetics
of transport and is essential to the stability of the steady state.

\smallskip
(ii) In the mesoscopic context, the scales for both elastic and
inelastic scattering become comparable to each other and to
the operative length of a conductor. Conventional bulk notions
of a purely geometric device ``length''; that is, one conceptually
divorced from the particular mechanisms for real scattering,
are problematic when device dimensions shrink dramatically
and its boundaries are wide open to direct energy exchange.

\smallskip
(iii) The current-fluctuation structure of a conductor is inseparable
from the current response. Transport models that are conserving always
lead to a fluctuation-dissipation relation, which will automatically
take care of the essential link between response and fluctuations.

\smallskip
(iv) Particularly in the meso- and nanoscopic regimes, strictly linear
models of current response may not be sufficient to discriminate among
subtle competing dynamical phenomena. For a fuller picture it would be
necessary to develop both nonlinear theoretical descriptions and
experiments that better revealed those properties lying beyond weak-field
measurement
\cite{gtd}.

\subsection{Many-body Aspects}

In strongly correlated systems, transport solutions that appear
straightforward in the semiclassical regime must be re-examined
in the appropriate many-body setting.
Of the Green-function techniques, the Keldysh analysis
\cite{langreth}
provides the paradigm of current interest
\cite{mw}.
The primary physical issues confronting a many-body approach remain
(1) whether it is inherently conserving; and (2) whether the FDT is
contained in it rather than having to be phenomenologically assumed.
A consistent Keldysh-based analysis will automatically meet both
requirements.

Since the first aim is to compute the single-particle current response,
one has to look for the particulars by which strong multi-particle
interactions filter down to the one-body level and modify the
noninteracting form of the conductance. The Kubo formalism already
identifies this nexus: it is the architecture of the correlation function
${\langle j(t) j(0) \rangle}$. In the Keldysh analysis, it is the
self-energy associated with the one-body propagator (Green function)
that ties the inner details of the many-body correlations to the outward
behaviour of the current response.

Briefly stated, a correlated system at rest possesses a self-energy
structure precisely attuned to the scattering amplitude
(interaction vertex) between carrier pairs. This tuning is essential
for microscopic conservation. It is encoded in the Ward identities
\cite{mahan}.

Scattering effects that appear in the form of a nonequilibrium
correction to the one-body self-energy (the so-called energy
relaxation rate) are systematically offset by counter-terms that
are nonequilibrium vertex parts, which are not of single-particle
form. However, there is a natural quantity that accommodates both
types: the two-particle propagator or correlation function,
characteristic of the Kubo formula.
In the semiclassical limit, the fundamental cancellations between
self-energy and vertex terms
manifest as the principle of detailed balance and determine
the conserving structure of the Boltzmann collision integral.

The Keldysh formalism lets one track the modified self-energy
structure in a controlled way. It can then be compared with the
equilibrium value. Taken in isolation, the self-energy difference
defines the loss rate in one-body propagation: the apparent exponential
attenuation in the strength of a single injected carrier reflecting
its progressive entanglement with innumerable many-body excitations.

Even though the computational accent is on the self-energy
we stress that, within the Keldysh expansion, the Ward identities
apply and require those contributions, not expressible as one-body
self-energy parts (namely the vertex corrections) to appear
explicitly in the overall relaxation. They do so on fully equal
terms with changes that keep the formal shape of a self-energy.
The complete form of the correlator ${\langle j(t) j(0) \rangle}$,
and thus (via Kubo and the FDT) the conductance, is the 
sum of two parts: self-energy and vertex terms.
Together, and only together, they determine the momentum relaxation rate
\cite{mahan},
whose inverse appears as $\tau$ in the Kubo formula,
Eq. (\ref{n15}).

A well-known application of Keldysh analysis is the Meir-Wingreen
formula
\cite{mw}

\begin{equation}
G = -G_0 \int dE
~{\rm Im} {\Bigl\{ {\langle \Gamma {\cal G}^r \rangle} \Bigr\}} 
{\partial \over \partial \mu} f^{\rm eq}(E - \mu)
\label{n30}
\end{equation}

\noindent
where the energy integral is over the asymptotic
carrier band of the (noninteracting) current reservoirs
attached to the central (interacting) system,
and the expression whose imaginary part is taken represents
the trace over the internal many-body excitations of the core.
The Green function ${\cal G}^r$ is in Keldysh retarded
form and $\Gamma$ encodes the inner core's
coupling to the leads.

In the noninteracting case,
$\Gamma(E, \omega) = 2\pi{\cal T}(\omega)\delta(E - \omega)$ is
the one-body transition-matrix operator
\cite{ziman2}
for the device treated as a barrier to single-particle propagation.
The corresponding free single-particle propagator is
${\cal G}^r(\omega, \omega')
= 1/(\omega - \omega' + i0^+)$
and the Meir-Wingreen kernel goes to

\begin{eqnarray}
\Gamma {\cal G}^r
&=&
\int {d\omega'\over 2\pi} \Gamma(E, \omega')
\int {d\omega\over 2\pi}
{e^{-i\omega\eta}\over \omega - \omega' + i0^+}
\cr
\cr
&=& -i{\cal T}(E);
~~\eta \to 0^+.
\label{mw1}
\end{eqnarray}

\noindent
Note again the need for an asymptotic convergence factor
to force this nondissipative analytic structure to simulate
a stable steady state, as explained in the preceding section. 
If ${\langle {\cal T}(E) \rangle}$ varies slowly with energy,
Eq. (\ref{n30}) recovers the Landauer formula Eq. (\ref{n24}).
In the interacting case, the integral on the
right side of Eq. (\ref{n30}) both formalizes and extends
Landauer's phenomenological transmission coefficient.

For a system thus described -- noninteracting or even interacting -- the
essential functions of dissipation and inelasticity appear to lack
clear definition in  Eq. (\ref{n30}). 
In that respect, three basic many-body aspects still await validation
within its expression.

\begin{itemize}
\item {\em Relation to Kubo formula and FDT}: how is the canonical
form of the two-particle Kubo correlation function recovered
from the integrand on the right-hand side of Eq. (\ref{n30})?

\item {\em Tracking the conserving structure of} $\Gamma$:
how does it conform to the Ward identities, dynamically
interconnecting the self-energy and vertex components?

\item {\em Auditing the mechanisms of energy loss}:
how do $\Gamma$ and ${\cal T}$ encode energy transfer (inelasticity)
from the driving field to the environment, via many-body
interaction(s)?
\end{itemize}

\section{Summary}

Mesoscopic electrical transport is now a major discipline within
condensed matter, and has been so for almost three decades.
Yet more than ever it is a rapidly changing, intellectually
stimulating and economically vital research area. In large part,
its theoretical initiatives are still fed by the astonishing growth,
in sophistication and versatility, of materials technology.

For all of those reasons it is important to hold to the fore, with
clarity, the conceptual underpinnings of transport theory. In
mesoscopics its points of reference are, and will remain,
precisely those that have governed the general understanding
of quantum kinetics: chiefly, the conservation laws and
the close interplay of electromagnetism with thermodynamics,
which leads to the universality of energy dissipation in real
mesoscopic systems.

The maturity of kinetic theory has been arrived at by a
collective effort going on, without pause, since the
formulation of quantum mechanics. Its legacy to mesoscopics
is a fruitful, almost extravagant arsenal of well crafted
and tested approaches to solving problems.
This applies not just to nonequilibrium mesoscopics but also
to the next step down, the quasi-atomic scale.

We have revisited the main, and already classic, kinetic techniques
with the goal of teasing out their common unifying strands.
These relate directly to the issues that will always
confront mesoscopic transport theory: microscopic conservation
and, most particularly, the accounting of energy dissipation
at small scales. If a theory wants to replicate their explicit
consequences, especially in a novel setting, it should first be able
to demonstrate that it explicitly respects their agency.

The dissipation issue led us to elucidate how the {\em nonlinear}
properties of energy loss processes hold the key to establishing
thermodynamic stability within a mesoscopic transport description.
The task of detailing in full the physics of dissipation outranks
even those pragmatic and increasingly pressing questions now
driving theoretical initiatives towards high fields.

Far from remote icons or symbolic figures for lip service, the
unifying ideas that we have traced here are the immediate and prime
constraints -- ``reality checks'' -- on every mesoscopic model aspiring
to serve the novel demands of device design by embodying, as practically
and flexibly as possible, deeply established physical rules.
We hope that this paper may help both as a reminder of them and as a
pointer to the questions that need to be addressed by mesoscopic
transport theories, before any answers are ventured.

\section*{Acknowledgment}

One of us (MPD) acknowledges the support of the Asia-Pacific
Centre for Theoretical Physics, Pohang, South Korea, where this work
was presented as part of the Focus Programme of September 2011.
In particular he thanks Professors Peter Fulde and Jongbae Hong
for their hospitality.


\begin{thebibliography}{99}
\bibitem{footnote4}
{Again, this is true
provided the system is normal and stable; phenomena such as
superconductivity and the fractional quantum Hall effect require
special measures to address the nature of their ground state,
before obtaining a description of the response, which always
involves excitations out of the ground state.}
\bibitem{footnote5}
{This is equivalent to attaching the additional
asymptotic convergence factor $e^{-\eta t}$ to the Kubo correlation function
$C(t)$, securing a stable limit within purely single-particle Hamiltonian
models; recall the discussion following Eq. (\ref{n14}).}
\end{thebibliography}
\end{document}